\documentclass[useAMS,usenatbib]{mn2e}

\usepackage{graphicx}
\usepackage{natbib}
\usepackage{booktabs}
\usepackage{subfigure}
\usepackage{ulem}

\def \aap {A\&A}
\def \aj {AJ}
\def \apj {ApJ}
\def \mnras {MNRAS}
\def \aaps {A\&AS}
\def \bain {Bul. Astron. Ins. Neth.}

\def \Teff {T_\mathrm{eff}}
\def \Msun {M_\odot}

\def \Lsun {L_\odot}

\title[TY\,CrA revisited]{The eclipsing binary TY\,CrA revisited: What near-IR light curves tell us
\thanks{Based on observations taken at the University Observatory of Bochum at 
Cerro Armazones, Chile. Based on observations made with the REM Telescope, 
INAF Chile under programmes 19002, 21024, and 23015.
Based on observations collected at the European Southern Observatory, Chile (programme
77.C-0549).
}}
\author[M. Va\v{n}ko et~al.]
{M. Va\v{n}ko$^{1}$\thanks{E-mail: vanko@ta3.sk}, 
 M. Ammler-von Eiff$^{2,3,4,5,6}$, T. Pribulla$^{1,3}$, R. Chini$^{7,8}$, 
 \newauthor E. Covino$^{9}$, R. Neuh\"auser$^{3}$\\
 $^1$Astronomical Institute of the Slovak Academy of Sciences, 059 60 Tatransk\'a Lomnica, Slovakia\\
 $^2$Th\"uringer Landessternwarte, Sternwarte 5, 07778 Tautenburg, Germany\\
 $^3$Astrophysikalisches Institut und Universit\"ats-Sternwarte, Schillerg\"a{\ss}chen 2-3, 07745 Jena, Germany\\
 $^4$Institut f\"ur Astrophysik, Friedrich-Hund-Platz 1, 37077 G\"ottingen, Germany\\
 $^5$Centro de Astronomia e Astrof\'isica da Universidade de Lisboa, Tapada da Ajuda, 1349-018 Lisboa, Portugal\\
 $^6$Centro de Astrof\'isica da Universidade do Porto, Rua das Estrelas, 4150-762 Porto, Portugal\\
 $^7$Astronomisches Institut, Ruhr-Universit\"at Bochum, Universit\"atsstra{\ss}e 150, D-44801 Bochum,
 Germany\\
 $^8$Instituto de Astronom\'ia, Universidad Cat\'olica del Norte, Antofagasta,
 Chile\\
 $^{9}$INAF -- Osservatorio Astronomico di Capodimonte, via Moiariello 16, 80131 Napoli, Italy}

\begin{document}

\date{Accepted 2012 December 15. Received 2012 November 01; in original form
2012 October 11}

\pagerange{\pageref{firstpage}--\pageref{lastpage}} \pubyear{2012}

\maketitle

\label{firstpage}

\begin{abstract}
New photometric observations of the hierarchical eclipsing TY\,CrA
system were taken in the optical with VYSOS6 and in the near-IR with
SOFI and REMIR. They are the first observations showing the deep eclipse
minimum of the pre-main sequence secondary in the near-IR. For the first
time, the secondary minimum can be reliably used in the calculation of
the O-C diagram of TY\,CrA. By now, the O-C diagram can be studied on a
time basis of about two decades. We confirm, that the O-C diagram cannot
be explained by the spectroscopic tertiary.
For the first time, the light curve of the inner eclipsing binary is
analysed in both optical and near-IR bands simultaneously. In
combination with already published spectroscopic elements, precise absolute 
dimensions and masses of the primary
and the secondary component are obtained using the ROCHE code. The
inclusion of the near-IR data puts strong constraints on the third light
which is composed of the reflection nebula, the spectroscopic tertiary
and a visual fourth component.
The absolute parameters of the inner eclipsing binary agree very well
with previous work except of the primary radius ($1.46\pm0.15$ $R_\odot$)
and luminosity ($40\pm10$ $L_\odot$) which are clearly smaller. While the
parameters of the secondary are well understood when assuming an age of
about 3-5\,Myrs, the primary seems considerably undersized. Low
metallicity cannot explain the parameters of the primary.
\end{abstract}


\section{Introduction}

% CrA star forming region:
The star-forming region R Coronae Australis (R~CrA) harbors dozens of
young intermediate- to low-mass stars (and brown dwarfs) of B8 to M8.5 spectral 
types. They have an age of between one and a few million years and a
distance of approximately 130 pc \citep[for a review see][]{ralph+08}. 
%The first few low-mass, 
%pre-main sequence stars, so-called T Tauri stars (TTS) in this star-forming region 
%were found by early H$\alpha $ and infrared (IR) imaging surveys 
%\citep{knacke+73, glass+75, marraco+81}.
%[Q1] Is the age range an estimate for the individual objects ?
%[N1] We might mention minimum mass for a star to set on the MS.

% TY CrA:
TY~CrA is a hierarchical triple, maybe even quadruple system, embedded in a
reflection nebula in the R Coronae Australis star forming region. The immediate environment of TY CrA 
has been studied recently in more detail by \citet{geers07}, \citet{juvela08}, \citet{currie11},
\citet{sandell11} and \citet{kumar11}. Two components 
form a massive eclipsing double-lined spectroscopic binary (eSB2) with an orbital 
period of almost 3 days. While the primary, a B8 star with a mass of
3.16~$M_{\odot}$, has already reached the main sequence, i.e. initiated stable hydrogen burning, 
the secondary still resides in the pre-main sequence (PMS) stage and has a mass of 1.64~$M_{\odot}$
\citep{casey+98}. A third spectroscopic component is in a wide orbit around the eclipsing 
pair \citep{casey+95, corporon+96}. A visual fourth component was detected by \citet{chauvin+03}.

% eclipsing SB2:
The analysis of eSB2 allows us to derive the most fundamental parameters, mass and radius, 
as well as to obtain effective temperature and luminosity.
%source: Casey et al. (1998, p. 1617/I 1)
Thus, the TY CrA eSB2 is one of the key objects that can provide precise constraints 
on the evolution of PMS stars. 
% Note: Can the orbital period of the

\begin{table*}
\caption{Parameters of the variable and comparison stars used. The
         proper motions and $(B-V) = 0.85(B-V)_T$ colors were taken from the
         Tycho-2 Catalog \citep{Tycho2}. The distance to the system TY~CrA
         was taken from \citet{chauvin+03} and infrared colors, $(J-K)$, 
         from 2MASS \citep{2MASS}; cmp5 = 2MASS~19013729-3649210 = 
         USNOB1 0531.0836104.  
\label{comparison_tab}}
% Do we give brightness difference of star #5 and HD176423 somewhere ?
% We should add star #5, RA & DEC of all stars !

\footnotesize
\begin{center}
\begin{tabular}{lccccccccl}
\hline
\hline
Comp.      &     GSC    &   HD    &    SAO   &  $\mu_\alpha$ [mas.y$^{-1}$] & $\mu_\delta$ [mas.y$^{-1}$] & $d$ [pc]           & $(B-V)$    & $(J-K)$   & sp. type \\
\hline
TY~CrA     & 7421-1126  &   --    &  210829  &  $-$0.2(1.3)                 & $-$28.6(1.2)                & 136$^{+25}_{-19}$  & 0.463(32)  & 0.813(23) & B8e      \\
cmp1       & 7421-1061  & 176423  &  210834  &  6.1(1.4)                    & $-$27.3(1.4)                &       --           & 0.291(20)  & 0.240(23) & A1III/IV \\
cmp3       & 7421-1163  & 176497  &  210840  &  3.3(1.3)                    & $-$28.5(1.3)                &       --           & 0.170(10)  & 0.080(28) & A0IV     \\
cmp5       &     --     &   --    &    --    &   --                         &  --                         &       --           &   --       & 1.939(33) & --       \\
\hline     
\end{tabular}
\end{center} 
\end{table*}

% effective temperature:
The derivation of effective temperature is a particularly important issue. 
%as effective temperature is the principal quantity in terms of stellar evolution. 
The analysis of the eclipse light curve (hereafter LC) yields the ratio of the effective 
temperatures of the two components but does not enable to determine individual
temperatures. Temperatures for both components can be derived if a reasonable 
estimate is used for one of them. This step is a major problem in the analysis of PMS 
stars since their effective temperatures are not well-known a priori
(\citealp{hillenbrand+04, ammler+05, schoning+08}). Because of a significant interstellar and circumstellar
absorption towards TY~CrA, observed colours have to be used with care \citep{casey+98}.
%[N2] The only reasoble way to determine Teff is to disentangle high S/N and high
% dispersion phase resolved spectra

% TY CrA: advantages for temperature determination:
TY\,CrA is one of only two known objects \citep[see][]{hillenbrand+04} which 
allow us to circumvent this problem: the primary has already reached the main sequence, 
so that its effective temperature can be estimated with a reasonable accuracy using the 
mass - temperature relation of main sequence stars. The secondary is still located 
in a PMS state.
%Q1: Is Teff=12000 K really derived from the evolutionary state of the primary ? 

% previous work: photometry & spectroscopy
The eclipsing nature of TY~CrA was reliably established by \citet{kardopolov+81} using
$BV$ photoelectric photometry. The authors observed $\sim$0.4 mag deep minima consistent
with an orbital period of 2.888777 days. Without spectroscopy it was, however, impossible to
exclude the double orbital period for the system. The first spectroscopic orbit
for TY~CrA was obtained by \citet{lagrange+93}. The authors showed that the radial-velocity 
(hereafter RV) variations of all narrow spectral lines are periodic and consistent
with the 2.888777 days period previously reported by \citet{kardopolov+81}. The width of lines, 
$v \sin i <$ 6 km~s$^{-1}$, indicated subsynchronous rotation. 
%Q3 possibly we should mention that it was the primary component !
The authors
noticed the second system of much wider lines anticorrelating 
in RV with respect to the narrow lines. \citet{corporon+94} confirmed that the broad
lines correspond to the secondary component and derived orbits for both components.
Later, \citet{casey+95} found that TY~CrA is a SB3 system. The RVs of the tertiary 
were found to correlate with the barycentric RV of the eclipsing pair
confirming the gravitational bond.  
The authors were able to derive a precise solution of the circular
orbit of the eclipsing binary after applying appropriate offsets to the primary
and secondary data at different epochs. The measurements were insufficient to derive a
unique solution of the tertiary orbit \citep[alse see][]{corporon+96} which most 
certainly has a long period compared to the orbit of the eclipsing binary.

% previous work: light curves
\cite{casey+98} obtained  {\it uvby} photoelectric photometry of TY~CrA \citep[see also][]{vaz+98}.
The measurements used HD176423, HD176497 as comparisons.
The authors confirmed the orbital period of the eclipsing binary of 2.89\,days. The primary with spectral 
type B8 has a mass of $3.16\pm0.02\,M_\odot$, the secondary $1.64\pm0.01\,M_\odot$. 
Following \citet{casey+98}, the age of the eclipsing binary is of the order of a few 
million years according to evolutionary models. The LC analysis was challenging 
mainly because of the very shallow secondary minimum in the optical bands with a depth 
of only $\approx0.02\,$mag. The authors claim that the accuracy of the LC is not limited 
by observational errors, which are below 0.01\,mag but by out-of-eclipse variability at a 
level of 0.05\,mag \citep{vaz+98}. The analysis is complicated by an offset of the 1992 
data relative to the 1993 data around phase 0.6 which indicates another type of variation, 
possibly due to stellar spots. \citet{casey+98} did not find a unique solution and pointed
out the need for more photometric data. In addition, there are effects due to the spectroscopic
tertiary (light-time effect), the third light detected in the LC analysis, 
the asymmetry of the primary minimum and the reflection nebula to be
considered.
%source: -no single most probable solution (Casey et al. 1998, p. 1627/II 2 
%[Q3] How important is LITE ? What would be its expected amplitude for the estimated
%period range ?

% multiplicity:
Observations of \citet{chauvin+03} with VLT/NACO showed a visual companion to TY~CrA separated by 0\farcs294. If
the additional component was a physical member of the system its spectral type would be M4. At such small separation the visual
component contributes to all ground-based photometry and spectroscopy. Its light contribution in NIR is about 4\% in 
the $J$, and 7.6\% in the $K$ passband, respectively. If physically bound, the fourth component would have significant 
consequences for the dynamical evolution of the system.
%Note: Because the proper motion of TY CrA in DEC is 28.6 mas/year additional AO imaging (10 years from above image)
%should easily solve the issue of the physical bond.
  
% new techniques:
The aim of this work is to improve the knowledge on TY~CrA and to solve open issues identified previously. 
% using NIR:
Recent work shows that uncertainties can be substantially reduced when involving observations in the near-infrared. 
\citet{covino+04} included near-infrared photometric data in the analysis of the young eclipsing binary 
RX~J0529.4+0041 and were able to reduce the error bars on mass and radius of the components by 
$\approx80\,\%$ compared to previous analyses in visual bands. The secondary minimum was deeper in the infrared band than 
in the optical thus allowing for higher accuracy. 
Also the determination of effective temperature from stellar colours benefits from the use of 
near-infrared observations, which are less affected by interstellar absorption.
%[N4] NA here. We have strong absorption + reflection nebulosity.
\citet{schoning+08} found that calibrations based on an optical pass band and an infrared pass band 
(e.g., $(V - K)$) give effective temperature for young stars which agree best with spectral types.

% present work:
The present work for the first time combines NIR light-curves (LCs) of TY\,CrA with observations in visual bands. 
The observational limitations encountered by \citet{vaz+98} in the visual are overcome by 
using the advantages of wide-field imaging where lots of comparison stars are exposed simultaneously 
together with the target. The goal of the present paper is to conclusively determine the effective 
temperature of the pre-main sequence secondary and to solve uncertainties in previous LCs solutions. 

\section{Observations and data reduction}

\subsection{Differential {\it {JH}} photometry}
\begin{table*}
\caption{Overview of telescopes/instruments and detectors used to obtain photometry 
         of TY CrA. Abbreviations of the observatories (Obs.): LS -- La Silla 
        (29\degr15\farcm0 S, 70\degr44\farcm0 W); CA -- Cerro Armazones (24\degr36\farcm0 S, 
         70\degr 11\farcm0 W); CT -- Cerro Tololo (30\degr10\farcm0 S, 70\degr48\farcm0 W). \label{instruments_tab}}
\footnotesize
\begin{center}
\begin{tabular}{llccccc}
\hline
\hline
Obs.             & Telescope/Instrument          &Aperture &Filters &Detector     & Size     &FoV\\
                 &                               &[cm]     &                                      &             &[pix]     &               \\
\hline
LS               & NTT/SOFI            &358      &$JHK_S$       & Hawaii HgCdTe   &1K$\times$1K & 4\farcm66 $\times$ 4\farcm66 \\
                 & REM/REMIR           &60       &$JHK_S$       &Hawaii I         &0.5K$\times$0.5K &10\farcm0 $\times$ 10\farcm0 \\ 
                 & REM/ROSS            &60       &$VRI$        &Apogee Alta      &1K$\times$1K  & 10\farcm0 $\times$ 10\farcm0     \\   
%                & 2.2m/FEROS          &220      &350-920nm  &EEV              &2k$\times$4K  &48,000           \\
CA               & VYSOS6              &15       &$BVRI$       &Apogee Alta U16M &4K$\times$4K  & 2\degr42\farcm0 $\times$ 2\degr42\farcm0 \\
%                & HPT/BESO            &150      &370-860nm  & E2V             &4K$\times$2K  &48,000         \\
CT               & 1.3m/ANDICAM        &130      &$BVRI$       &Fairchild 447    &2K$\times$2K  & 6\farcm0 $\times$ 6\farcm0           \\
                 &                     &130      &$YJHK$       &Hawaii HgCdTe    &1K$\times$1K  & 2\farcm4 $\times$ 2\farcm4        \\
\hline
\hline
\end{tabular}
\end{center}
\end{table*}
%[N5] Is it necessary to write HgCdTe ? All detectors covering the K band use those elements...

%SOFI
First {\it JHK$_\mathrm{s}$} observations for this project were taken during two nights (July 30 
and September 7) in 2006 with the SOFI instrument \citep[see][and Table~\ref{instruments_tab}]{moorwood+98} at the 3.5m ESO 
NTT at La Silla, Chile. The observations were scheduled to cover one secondary and one primary minimum of the
system.

The observations during the first night suffered from strongly variable conditions and were not used in the
analysis. The primary minimum observed in the second night is of good quality. In addition to precise minima
timing (see Section 3), the data were used for the LC analysis. All SOFI observations use cmp5 as the comparison
star (see Table~\ref{comparison_tab} and Figure~\ref{vis_ir})

\begin{figure}
\centering
\includegraphics[width=8cm]{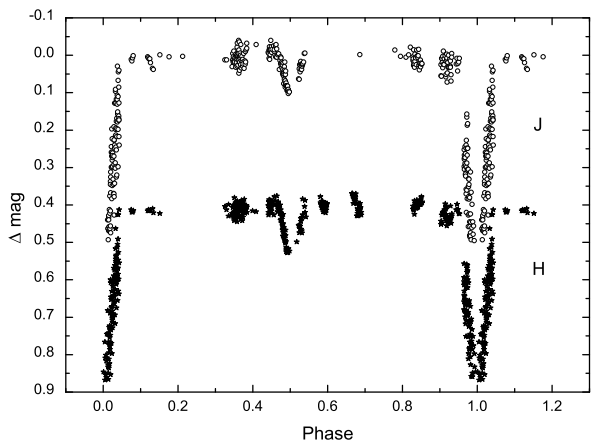}
\caption{\label{jh_fig} The $JH$ LC of TY~CrA obtained by us with REMIR (2009 and 2011) and SOFI (2006) 
at La Silla. The LC is incomplete because of bad weather and pointing problem. The primary minimum was 
obtained with SOFI in 2006 and the rest of LC was observed with REMIR in 2009 and 2011. The
observations from 2011 were reduced with respect to HD\,176423. For other observations we used cmp5 as 
the comparison star (see Table~\ref{comparison_tab}). Therefore, to obtain whole LC we took into account the
brightness difference of the comparison stars (see the text). Note that the errors of
LC points in $JH$ photometry are within the range of 0.003 -- 0.010 mag.
}
\end{figure}

%REM:
Additional infrared photometry was taken in 2009-2011 with REM\footnote{http://www.rem.inaf.it/} \citep[Rapid Eye Mount,][] {zerbi+01, 
antonelli+03, zerbi+04} (60cm diameter rapid response telescope) operated at La Silla, Chile. The telescope 
hosts two instruments: REMIR, an infrared imaging camera, and ROSS, a 
visible light imager. The two cameras can observe simultaneously thanks to a 
dichroic plate placed before the telescope focus. %the same field.

In order to fully cover the NIR {\it JHK$_\mathrm{s}$} LC we got REM observing time by 
three proposals (in the periods AOT19, AOT21 and AOT23). The observations strongly suffered from bad weather 
and technical problems. The observing programme has been 
completed to a low percentage even though additional DDT time was granted. The first observations taken in AOT19 
and 21 followed the standard approach of differential photometry, i.e. a pre-defined field is observed containing
the target and the selected comparison star (cmp5) (see Figure~\ref{vis_ir}). 
However, the REM frames are affected by pointing and off-centering 
problems, so that comparison star and sometimes also the target were off the frames.
% Maybe we should note there are few photometrically stable stars around TY CrA...
The 2010 data (AOT21) are of low quality and were excluded from the LC analysis.
For period AOT23, we followed a different observing strategy. Since the pointing was not sufficiently precise, 
we decided to observe other fields including comparison stars already used by \citet{casey+98} and \citet{vaz+98}, and
to center the pointing at these stars (HD\,176423 (cmp1) and HD\,176497 (cmp3)).
These fields and TY\,CrA were observed in an alternating sequence with high frequency which must not be lower 
than the frequency of variations of the near-infrared background. The observational strategy was further 
improved by observing selected critical phases (ingress, egress, minimum) of many eclipses distributed over
different nights instead of observing only a few eclipses continuously from ingress to egress. This way, 
the loss of a single night due to weather or technical problems had less effect on the overall success of 
the observations. Eventually, we obtained 2 secondary minima and out-of-eclipse observations around phases 0.25 and
0.75.
% Did we use the same comparison star for all observing runs ?

%ANDICAM
The same observing strategy was used also in additional near-IR photometry carried out by ANDICAM. This is a 
simultaneous optical and NIR imaging instrument \citep{depoy+03} operated by the SMARTS consortium at 
the 1.3m CTIO telescope, Chile. The field of view of the NIR camera is roughly centered on the larger field 
of view of the optical arm. We observed in two nights in May 2011; however, because of bad observing conditions 
(we obtained only several usable datapoints), this observation was excluded from the following analysis.

Concerning comparison stars, note that, in the case of SOFI data (2006) and REM data from  2009, we used 
2MASS~19013729-3649210 (RA$_{2000}$: 19:01:37.292, DEC$_{2000}$:-36:49:21.04) as additional comparison star.
The REM observations in 2011 were measured with respect to HD\,176423. The additional standard 
is fainter than HD\,176423 in the $J$ and $H$ bands, by 
1.799 $\pm$ 0.006 and 0.642 $\pm$ 0.004  respectively. 
The $K_S$ band observations, especially in the case of REM, were very scattered. Hence, for the photometric 
analysis we could use only the $JH$ LCs.

%data reduction:
The near-IR {\it JH} data reduction was implemented using IDL\footnote{http://idlastro.gsfc.nasa.gov/contents.html}, 
ESO MIDAS\footnote{http://www.eso.org/sci/software/esomidas/}, and ESO ECLIPSE\footnote{http://www.eso.org/sci/software/eclipse/}. 
The flat-field frames and the bad pixel masks were created using the ESO ECLIPSE {\textit
{flat}} and {\textit {average}} recipes.
The flat-field division, bad pixel correction, sky subtraction, and shift and add of the science frames was performed 
using the ESO ECLIPSE {\textit {jitter}} recipe.\footnote{Each REMIR jitter sequence consists of five frames. The jitter pattern has a 
pentagon shape and is exactly the same for each sequence but might be rotated. Mostly, the offsets were made available to the 
recipe beforehand and refined by cross-correlation search when possible. In many cases, however, the {\textit
{jitter}} recipe even succeeded in performing a blind offset search.} 

In further reduction tasks, the IRAF package\footnote{http://iraf.noao.edu/} has been used. In the next step the astrometric 
solution of the frames was performed. The WCS system was determined using the 2MASS on-line catalogue for 
reference\footnote{http://www.ipac.caltech.edu/2mass/releases/allsky/}. 
Then the aperture photometry of the target and comparison stars has been done. The instrumental magnitudes for several apertures appropriate 
for the seeing conditions at site were determined. The details of all near-IR observations are mentioned in the journal 
of observations in the Table~\ref{journal_tab}. The differential extinction was neglected.
%Maybe we should estimate its magnitude...

\begin{figure}
\centering
\includegraphics[width=8cm]{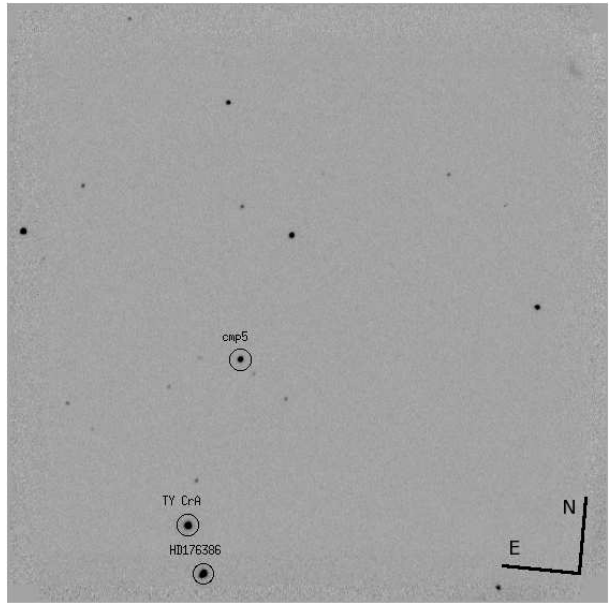}
\caption{\label{vis_ir} The field around TY\,CrA in the H passband (the
image taken on June 27, 2009).}
\end{figure}

Note that the errors of the LC data points in $\it JH$ photometry are between 0.003 and 0.010 mag. 
In contrast to the optical images, there is no contamination due to the reflection nebula. 
Figure~\ref{vis_ir} shows that in the near-IR there is no background emission
from the nebula, contrarily to the optical where a strong contribution
from the reflection nebula is observed.

Interestingly, we noticed an asymmetry in the primary minimum in the near-IR SOFI data which 
was already noticed by \citet{casey+98} in the optical.

\subsection{{\it BVRI} CCD photometry}

%VYSOS6:
The optical CCD photometry of TY CrA was performed by VYSOS6, 
a wide field imaging instrument located at Cerro Armazones 
Observatory, Chile
\citep{haas+11}. 

\begin{figure}
\centering
\includegraphics[width=8cm]{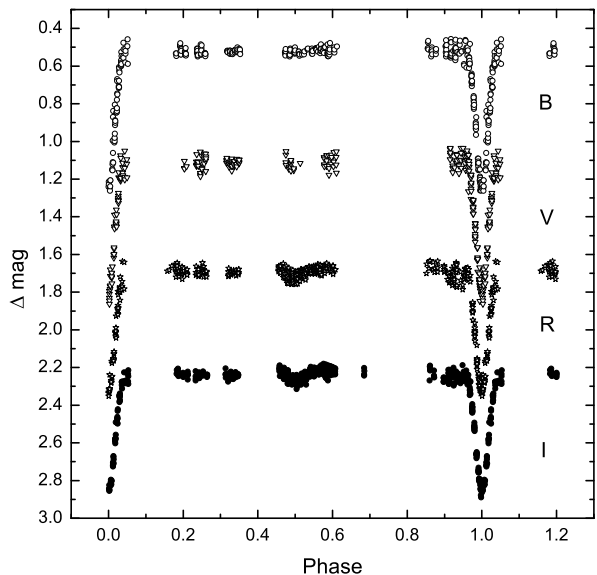}
\caption{\label{bvri_fig} The {\it BVRI} LC obtained
by us with VYSOS6 at the Cerro Armazones in 2009. The errors of individual points
in the {\it BVRI} LCs are in the range of 0.006--0.009 mag. The LCs were arbitrarily
shifted in magnitudes for clarity.}
\end{figure}

\begin{figure}
\centering
\includegraphics[width=8cm]{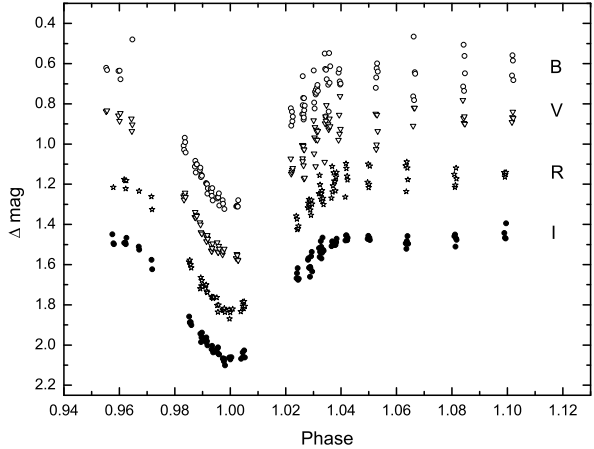}
\caption{\label{primary_fig} Primary minimum in {\it BVRI} obtained
with VYSOS6 at the Cerro Armazones observatory in 2011.}
\end{figure}

\begin{table}
\caption{The journal of CCD ({\it BVRIJH}) observations of
TY CrA obtained at La Silla (REMIR, SOFI) and Cerro Armazones (VYSOS6)  
observatories in the period 2006 -- 2011.
The julian date HJD$_{mean}$ is corresponding to the beginning of
observation.
N$_{im.}$ -- number of CCD frames in $\it R$ and $\it J$ filter.   
\label{journal_tab}} 
\footnotesize  
\begin{center} 
\begin{tabular}{ccccc}
\hline
\hline
Date         & HJD$_{mean}$    &     Phase       & Filter & N$_{im.}$\\
             & 2\,400\,000+   &                 &        &             \\
\hline
VYSOS6       &                 &                 &        &             \\
\hline 
Jun 26, 09  & 55009.719  & 0.184 -- 0.211  &$ BVRI $   & 24 \\
Jun 27, 09  & 55010.565  & 0.476 -- 0.593  &$ BVRI $   & 84 \\ 
Jun 28, 09  & 55011.786  & 0.899 -- 0.933  &$ BVRI $   & 25  \\
Jul 22, 09  & 55035.644  & 0.158 -- 0.181  &$ BVRI $   & 8  \\ 
Jul 23, 09  & 55036.505  & 0.456 -- 0.565  &$ BVRI $   & 98 \\ 
Jul 24, 09  & 55037.648  & 0.852 -- 0.882  &$ BVRI $   & 20  \\
Aug 26, 09  & 55070.530  & 0.234 -- 0.263  &$ BVRI $   & 23  \\
Aug 27, 09  & 55071.518  & 0.576 -- 0.609  &$ BVRI $   & 17  \\
Aug 28, 09  & 55072.492  & 0.914 -- 0.993  &$ BVRI $   & 44 \\ 
Aug 31, 09  & 55075.494  & 0.953 -- 0.031  &$ BVRI $   & 59  \\
Sep 01, 09  & 55076.559  & 0.321 -- 0.349  &$ BVRI $   & 25 \\ 
Sep 02, 09  & 55078.496  & 0.992 -- 0.048  &$ BVRI $   & 41 \\ 
May 27, 11  & 55709.812  & 0.699 -- 0.711  &$ BVRI $   & 20\\  
May 29, 11  & 55711.851  & 0.406 -- 0.420  &$ BVRI $   & 20\\  
May 30, 11  & 55712.674  & 0.690 -- 0.703  &$ BVRI $   & 19\\  
Jun 04, 11  & 55717.879  & 0.492 -- 0.506  &$ BVRI $   & 20\\  
Jun 06, 11  & 55719.707  & 0.125 -- 0.172  &$ BVRI $   & 27\\  
Jun 08, 11  & 55721.826  & 0.858 -- 0.874  &$ BVRI $   & 21\\  
Jun 09, 11  & 55722.674  & 0.152 -- 0.209  &$ BVRI $   & 40\\  
Jun 10, 11  & 55723.681  & 0.501 -- 0.514  &$ BVRI $   & 21\\  
Jun 11, 11  & 55724.809  & 0.891 -- 0.906  &$ BVRI $   & 20\\  
Jun 12, 11  & 55725.893  & 0.191 -- 0.266  &$ BVRI $   & 29\\
\hline
\hline
REM-IR      &             &                &           &   \\
\hline
\hline
Apr 02, 09 & 54923.907  & 0.478 -- 0.481  &$ JH $ & 5 \\
Jun 13, 09 & 54995.740  & 0.344 -- 0.368  &$ JH $ & 29\\
Jul 23, 09 & 55036.469  & 0.443 -- 0.456  &$ JH $ & 12\\
Jul 24, 09 & 55036.509  & 0.457 -- 0.489  &$ JH $ & 39\\
Jul 25, 09 & 55037.801  & 0.344 -- 0.368  &$ JH $ & 29\\
Aug 20, 09 & 55063.585  & 0.830 -- 0.852  &$ JH $ & 28\\
Sep 05, 09 & 55079.567  & 0.363 -- 0.384  &$ JH $ & 24\\
Jun 12, 09 & 55359.749  & 0.352 -- 0.380  &$ JH $ & 24\\
Jun 22, 09 & 55369.732  & 0.808 -- 0.832  &$ JH $ & 21\\
Apr 25, 11 & 55676.713  & 0.074 -- 0.126  &$ JH $ & 9\\ 
May 01, 11 & 55682.712  & 0.150 -- 0.494  &$ JH $ & 6\\ 
Jun 23, 11 & 55735.680  & 0.486 -- 0.539  &$ JH $ & 12\\
Aug 14, 11 & 55787.552  & 0.442 -- 0.527  &$ JH $ & 4\\ 
\hline
\hline
SOFI       &&&&\\
\hline
\hline
Sep 07, 06 & 53986.460  & 0.964 -- 0.042 & $ JH $ & 242\\
\hline
\hline
\end{tabular}
\end{center} 
\end{table}  

Observations of TY\,CrA were taken in 2009 and 2011 (Fig.~\ref{bvri_fig}). By the second campaign
in 2011, VYSOS6 got an identical twin installed on the same mount.
One telescope offers the filter combination $\it VR$ and the other $\it BI$,
so that observations are taken in two filters simultaneously.
Both telescopes are equipped with an Apogee Alta U16M 4096$\times$4096 
pixels CCD camera each, providing a 2\degr42\farcm0 $\times$ 2\degr42\farcm0 field of
view. During the first observing campaign we have covered the whole LC in    
{\it BVRI}, used for the analysis described in the following. Unfortunately, in the case of the 2011 photometry,
we obtained only an incomplete primary minimum (see Fig.~\ref{primary_fig}). The secondary
minimum was not observed at all because of bad weather.

For the optical differential photometry, we also used HD~176423 and HD~176497 
as comparison and check star, respectively. The stars were found to be constant 
within the observational accuracy throughout the observing period. 
We reduced all optical data in a standard manner (bias, dark, flat-field
corrections), and aperture photometry was performed using IRAF and custom 
written tools. 

The LCs obtained in 2009 with VYSOS6 show that the depth of the 
secondary minimum (at phase 0.5) increases towards the redder
bands ({\it R}, {\it I}) and reaches almost 0.10~mag in the $I$-band.
The errors of individual points in the {\it BVRI} LCs are in the range of
0.006--0.009 mag. 

The visual LCs of TY~CrA are contaminated by the reflection nebula. Part of its
light is subtracted as background within the aperture photometry. Any remaining contribution 
is corrected for by fitting the third light in the LC analysis.

%Note: we should specify the apertures used in the visual and NIR photometry !!!
%Q: Do we see enhanced scatter of TY CrA compared to other stars ???

\section{Minima determination and the light-time effect}
\label{minima}

Our observations enabled us to determine 3 minima times of the TY~CrA eclipsing pair\footnote{Partially 
covered eclipses were not used for the timing.}, which add to the \citet{casey+98} data and one
recent observation  by \citet{paschke10}. They are listed in Tab.~\ref{minima_tab}. The times 
of minima were determined separately for all filters using the \citet{kwee+56} method and then the 
weighted averages were calculated.

For computation of minima times we have used only data in the phase interval $\pm$0.05 around either of the minima. 
This approach minimizes the influence of the minima asymmetries \citep{veer73}. 

\begin{table}
\caption{Primary and secondary minima times of TY~CrA. For some minima, the filters used were not given in the
original publication \citep{casey+98}. For new minima observed in more than one filter the weighted
average is given.
\label{minima_tab}}
\footnotesize
\begin{center}
\begin{tabular}{lccc}
\hline
\hline
HJD        & $\sigma$ & Filter &  Type   \\
2\,400\,000+  & [days]   &          &         \\
\hline
\citet{casey+98} &&& \\
\hline
47694.7971 & 0.0009 & -- &      pri \\
47700.5740 & 0.0005 & -- &      pri \\
47710.68   & 0.08   & -- &      sec \\
48783.8703 & 0.0013 & -- &      pri \\
49153.6302 & 0.0032 & $y$&      pri \\
49160.85   & 0.02   & $y$&      sec \\
49163.74   & 0.01   & $yb$ &    sec \\
49514.7310 & 0.0003 & -- &      pri \\
49527.738  & 0.005  & -- &      sec \\
49537.8414 & 0.0008 & -- &      pri \\
49543.6193 & 0.0003 & -- &      pri \\
\hline
This work  &&&\\
\hline
53986.55284 & 0.00038 & $JH$  & pri \\
55036.64046 & 0.00035 & $IR$  & sec \\
55075.63231 & 0.00016 & $BVRI$& pri \\
\hline
\citet{paschke10} &&&\\
\hline
55387.619  & 0.005    & $V$   & pri \\
\hline
\hline
\end{tabular}
\end{center}
\end{table}

\begin{figure}
\centering
\includegraphics[width=8cm]{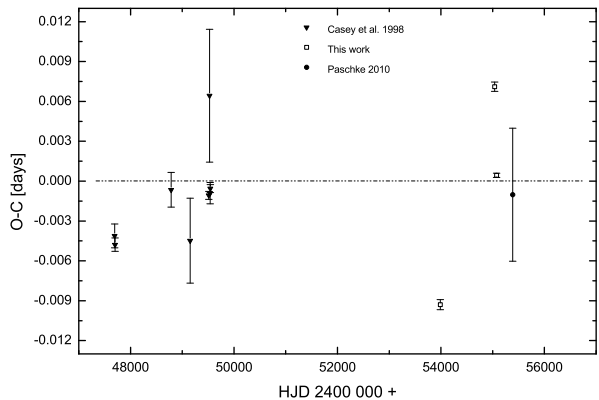}
\caption{\label{omc_fig} (O-C) diagram for all published and new minima (obtained from optical and
NIR LCs) timings corresponding to the optimal ephemeris obtained by the simultaneous multi-colour 
LC analysis (Table~\ref{elements_tab}).
}
\end{figure}

Figure~\ref{omc_fig} shows the (O-C) residuals for all available minima (Table~\ref{minima_tab}) with respect
to the optimum ephemeris determined by the LC modeling (Table~\ref{elements_tab}). As expected, for a detached
binary, the orbital period of the system is stable. Nevertheless, the (O-C) residuals exhibit 
deviations significantly larger than their standard errors. \citet{corporon+96} presented five different 
spectroscopic orbital solutions for periods of the outer orbit $P_3$ = 126-270 days. The most probable are their 
solutions \#1 ($P_3$ = 270 days) and \#2 ($P_3$ = 268 days). The expected light-time effect (hereafter LITE) 
amplitude modulating the binary eclipse timings was computed for solution \#1 and the following masses of the 
components: $M_1$ = 3 $M_\odot$, $M_2$ = 1.6 $M_\odot$ and $M_3$ = 1.2 $M_\odot$. Using the published major
axis $a$ = 1.47 a.u., the outer orbit inclination $i$ = 20$^\circ$, and the above masses we get the projected
semi-major axis of the eclipsing pair as: $a_{12} \sin i$ = 0.104 a.u. corresponding to 51.9 light seconds
(or 0.0006 days). In the view of the most precise minima having $\sigma$ = 0.0002 days and probable systematic
errors (caused e.g. by activity of TY~CrA), the detection and useful analysis of the timing signal caused by 
the third component can hardly be performed using the ground-based photometry.

Because of the short period of the outer orbital pair the RV observations are more sensitive and useful in the
decisive determination of the orbital elements. To summarize, we agree with \citet{casey+98} 
in that the photometric O-C diagram of the inner eclipsing binary cannot be explained by orbital solutions  
for the spectroscopic tertiary suggested previously by \citet{corporon+96}.

\section{Light curve analysis}
%light curve in several bands:

Unfortunately, optical ({\it BVRI}) data and near-IR ({\it JH}) photometry were obtained in different
time intervals (see Table~\ref{journal_tab}). Most of the optical photometry was taken in 2009, 
while the near infrared data used for analysis in 2006, 2009, and 2011. Because of the extreme stability of LCs, we 
%Note: Is that true ?
attempted to model all photometric data simultaneously. Separate solutions of visual and NIR data were 
also performed. 

Our photometry of TY~CrA is the sum of the eclipse LC of the inner binary and a constant light 
contribution of the unresolved third component and the visual companion (0.3 arcsec from the unresolved 
triple). Moreover, TY CrA and HD~176386 (1\farcm1 southwest of TY CrA) illuminate the reflection nebula NGC~6726/6727, 
whose light is a significant contributor to the observed flux \citep{casey+98}. The light from the two 
stars in the eclipsing pair is of principal interest here. The remaining sources produce a constant offset, so-called 
third light, decreasing the LC amplitude and complicating the analysis. Because of the three 
different sources, the third light cannot be modeled as a single stellar source (defined by its
fractional radius, and effective temperature). Hence the third light will be treated without any
assumptions and independently in the individual passbands.

The latest version of the code {\it ROCHE} was employed \citep{pribulla+12}.
The LC of the system was modeled assuming (1) radiative envelopes with bolometric albedos $A_1 = A_2$ = 1.00 
\citep[e.g.][]{rucinski69} and gravity darkening coefficients $g_1 = g_2 = 1.00$ appropriate for the 
B8V spectral type \citep[see][]{pantazis+98}, 
% What about the secondary component ? Does it still have radiative envelope ?
(2) limb darkening coefficients automatically interpolated from extensive tables of \citet{vanhamme93} for 
average surface gravity $\log g$, average $T_{eff}$ and given passband, (3) solid 
body rotation of the components (no differential rotation), (4) synchronous rotation of the secondary component 
$F_2$ = 1 and no rotation of the primary $F_1$ = 0 (see below), (5) rotational axes perpendicular to the orbital
plane, (6) circular orbit, (7) no photospheric spot(s), and (8) local intensity computed using model atmospheres 
(local $\log g$, $T_{eff}$, and $\lambda$) for the solar chemical composition. The mutual irradiation 
effects were approximated by single reflection. Surface grids were derived from a icosahedron in such a way that 
surface elements are close to equilateral triangles of similar size. 

In the optimization process the following parameters were adjusted: $T_0, P$ - corresponding to the 
primary minimum and orbital period, $i$ - inclination angle, $\Omega_1, \Omega_2$ - generalized surface 
equipotentials, $T_2$ mean temperature of the less massive and colder secondary component, 
$l_1 (\lambda)+ l_2 (\lambda)$ - passband-specific unit-less normalization factors, and 
$l_3 (\lambda)$ - third light\footnote{The third light is expressed in ROCHE as $l_3 = L_3/(L_1 + L_2)$.}. 
The temperature of the primary $T_1$~=~12000$\pm$500~K was adopted from \citet{casey+98} while
the spectroscopic orbit ($K_1 = 85.2\pm0.2$ km.s$^{-1}$, $K_2 = 164.6\pm1.6$ km.s$^{-1}$) was adopted 
(not adjusted) from \citet{casey+95}. 

Before arriving at the final solution we extensively tested the subsynchronous rotation hypothesis. Because
of the small fractional radius\footnote{relative to the major axis $a$, unitless quantity} of the primary 
component ($r_2 \sim 0.1$) the asynchronism has
a very small effect on the LC shape. The relative change of the equatorial radius of the primary component
between the case of no rotation ($F_1$ = 0) and synchronous rotation ($F_1$ = 1) is about 0.8\% only.
Hence we assume that the primary does not rotate at all. The multi-colour LCs were solved by the method of differential
corrections. The optimisation process always resulted in the same parameter set (we used
different starting parameters) indicating uniqueness of the solution. 
% The fractional radii are given relative to the orbital semi-major axis.
%These third-light parameters were then adopted and held fixed while the 
%remaining photometric elements were adjusted during the following LC solution runs.

\begin{figure}
\centering
\includegraphics[width=8cm]{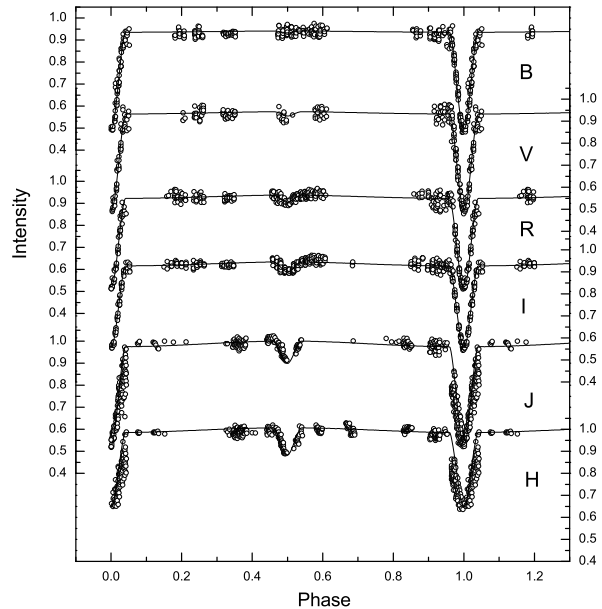}
\caption{\label{lcfits_fig} Best fits to the observed {\it BVRIJH} LCs of TY CrA resulting from the simultaneous modeling 
of all bands. The orbital phases are determined according to the new ephemeris given in 
Table~\ref{elements_tab}.}
\end{figure}

The derived photometric elements are presented in Table~\ref{elements_tab}. The values found for $l_3$ 
are 0.8965, 0.6969, 0.6791, 0.6940, 0.3032, and 0.4205 for $B$, $V$, $R$, $I$, $J$ and $H$, respectively, 
showing the increasing influence of the nebular contribution for shorter wavelength (see Table~\ref{elements_tab}), 
the additional stellar component(s) dominating in the NIR. The best fit to the {\it BVRIJH} LCs is displayed in 
Fig.~\ref{lcfits_fig}. In the case of the $H$-band data the primary minimum together with fit has lower 
($\sim$ 0.08~mag) amplitude than in the other filters. This fact is due to the bad data quality in 
the primary minimum. The scattered points have been removed so that the amplitude of the fit is lower 
and the third light contribution is larger. The {\it BVRIJH} LCs solution for synchronous rotation and rotational axes 
perpendicular to the orbital plane predicts $v_1 \sin i$ = 25.4 km~s$^{-1}$ and 
$v_2 \sin i$ = 40.0 km~s$^{-1}$.

% Note: we must be careful with the increasing contribution of the nebulosity: having two more stellar objects
% + different apertures in the visual in NIR it is complicated.

The absolute parameters of the primary and secondary components of TY CrA (corresponding to the simultaneous fit to 
the ({\it BVRIJH}) LCs) are given in Table~\ref{absolute_tab}. The values of most parameters, 
except of $R_1$ and $L_1$, are consistent with those derived by \citet{casey+98}. We obtained about 19\% smaller 
diameter of primary and about 10\% larger diameter of the secondary component compared to values given by 
\citet{casey+98}.
Such a difference in diameter translates to even larger difference in luminosity: our value for the primary is by 42\% lower. 
This discrepancy will be discussed in the following section.

To better compare our results and those of \citet{casey+98}, we also performed an analysis of the optical 
{\it BVRI} LCs only. The resulting parameters given in Table~\ref{absolute_tab} are still not 
consistent with those presented by \citet{casey+98}: $R_{1}$ and $L_{1}$ are about 15\%  
and 35\% smaller, respectively.  
      
In addition, we solved the LCs in the $BVRI$ passbands without third light reproducing very closely the 
parameters given by \citet{casey+98}. However a good solution in the $J$ and $H$ bands cannot be 
achieved without admitting a third light contribution.  The conclusion is that the near-IR data 
allows us to constrain the third light very well. 
      
\begin{table}   
\caption{Photometric elements (coming from {\it BVRI} and {\it JH} simultaneous analysis) 
of the eclipsing pair in the TY~CrA system  and their standard errors, $\sigma$ (number in
parentheses gives the error of the last digit) -- $i$ -- inclination; $e$ --
eccentricity; $q$ = $m_2/m_1$ -- mass ratio and $K_1 + K_2$ -- sum of radial-velocity semi-amplitudes 
\citep[adopted from][]{casey+95}; $\Omega_{1,2}$ -- generalized surface 
potentials; $f_{1,2}$ -- asynchronous rotation factors, $T_2$ -- effective temperature of the secondary 
component; $r_{pri}$, $r_{sec}$ -- volume mean fractional radii; $L_1^{B}/(L_1^{B}+L_2^{B})$ -- 
monochromatic lights and $l_3$ -- contribution of the third light. Parameters fixed (not adjusted) 
in the solution are denoted by a superscript "$f$". The uncertainty of the secondary-component
temperature contains quadratically added 200 K error resulting from the 500 K error of the primary-component
temperature estimated by \citet{casey+98}. The error has not been propagated to uncertainties of passband-specific
relative lights $L_1/(L_1+L_2)$.
\label{elements_tab}}
\footnotesize 
\begin{center}
\begin{tabular}{lc}
\hline
\hline
Parameter     & TY CrA         \\
\hline
HJD$_0$       & 2\,449\,537.8423(25) \\
$P$ [days]    & 2.88877912(21) \\
$i$ [$\degr$] & 85.1(10)       \\
$e$           &  0$^f$         \\
$\Omega_1$    & 10.4(10)       \\
$\Omega_2$    & 4.5(6)         \\
$F_1$         & 0.0$^f$        \\
$F_2$         & 1.0$^f$        \\
$q$           & 0.518$^f$      \\
$K_1 + K_2$ [km.s$^{-1}$]  & 249.8$^f$ \\
$T_1$ [K]     & 12000$^f$      \\
$T_2$ [K]     & 4900(250)      \\
$r_{pri}$     & 0.102(10)      \\
$r_{sec}$     & 0.161(9)       \\
\hline
$L_1^{B}/(L_1^{B}+L_2^{B})$ & 0.9674(14)\\
$L_1^{V}/(L_1^{V}+L_2^{V})$ & 0.9177(32)\\
$L_1^{R}/(L_1^{R}+L_2^{R})$ & 0.8581(10)\\
$L_1^{I}/(L_1^{I}+L_2^{I})$ & 0.7905(9)\\
$L_1^{J}/(L_1^{J}+L_2^{J})$ & 0.6592(13)\\
$L_1^{H}/(L_1^{H}+L_2^{H})$ & 0.5400(11)\\
\hline
$l{_3^B}$    &  0.8965    \\
$l{_3^V}$    &  0.6969    \\
$l{_3^R}$    &  0.6791    \\
$l{_3^I}$    &  0.6940    \\
$l{_3^J}$    &  0.3032    \\
$l{_3^H}$    &  0.4205    \\
\hline
\hline
\end{tabular}
\end{center}
\end{table}
%Note: we should add the standard errors of third light

\begin{table}   
\caption{Absolute parameters of the TY~CrA eclipsing pair. The second column corresponds to solution \#1 from the analysis 
of {\it BVRIJH} LCs. The third column shows solution \#2 resulting from analysis of {\it BVRI} LCs
only. {The mass ratio, $q$ and $K_1 + K_2$ were adopted from the circular solution of} \citet{casey+95}. 
All results are compared with values published by \citet{casey+98}. The larger masses for solution 2
result from lower inclination angle, $i$ = 82.9$\degr$.   
\label{absolute_tab}}
\footnotesize 
\begin{center}
\begin{tabular}{lccc}
\hline
\hline
Param.             & Solut.~1          &  Solut.~2       &  Casey et al.  \\
                   & [$BVRIJH$]        & [$BVRI$]        &      (1998)    \\
\hline
$M_1$/$M_\odot$    &  3.11(7)    & 3.15(7)     & 3.16(2)       \\ 
$M_2$/$M_\odot$    &  1.61(2)    & 1.63(2)     & 1.64(1)       \\
$R_1$/$R_\odot$    &  1.46(15)   & 1.53(13)    & 1.80(10)      \\
$R_2$/$R_\odot$    &  2.30(14)   & 2.21(12)    & 2.08(14)      \\
$T_1$ [K]          &  12000(500) & 12000(500)  & 12000(500)    \\
$T_2$ [K]          &  4894(250)  &  4977(450)  &  4900(400)    \\
$L_1$/$L_\odot$    &  40(10)     &     44(10)  &    67(12)     \\
$L_2$/$L_\odot$    &  2.7(7)     &    2.7(10)  &   2.4(8)      \\
$\log~g_{1}$ [cgs] &  4.60(8)    &   4.57(7)   &  4.43(6)      \\
$\log~g_{2}$ [cgs] &  3.92(5)    &   3.96(4)   &  4.02(5)      \\
$a$ [$R_\odot$]    &  14.29(9)   &  14.35(9)   &  14.29(9)     \\
\hline
\hline
\end{tabular}
\end{center}
\end{table}

\section{The evolutionary state of TY~CrA}
\subsection{Mass-radius diagram}

\begin{figure}
% \vspace*{-2.0 cm}
\begin{center}
 \includegraphics[width=8.2cm]{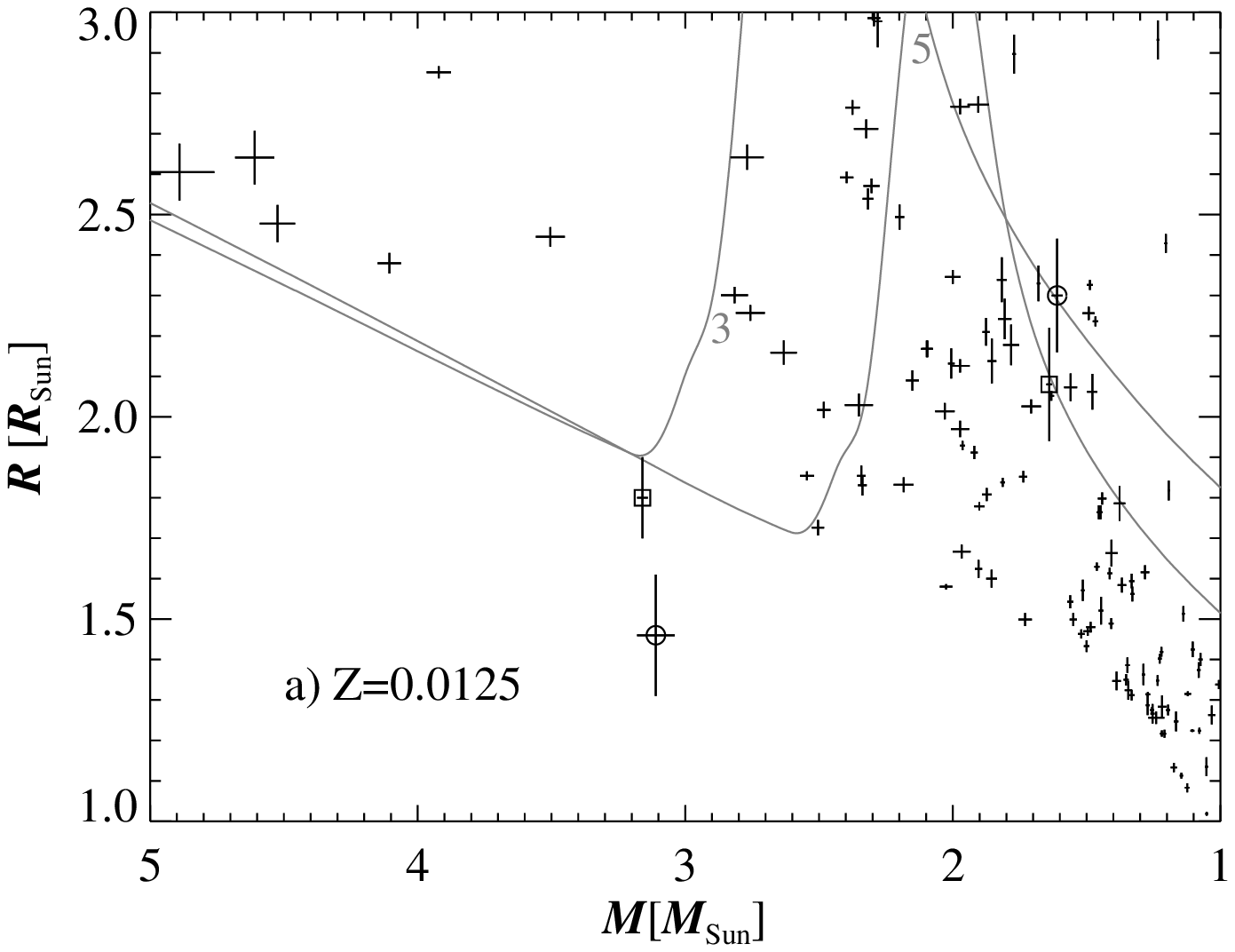}
 \includegraphics[width=8.2cm]{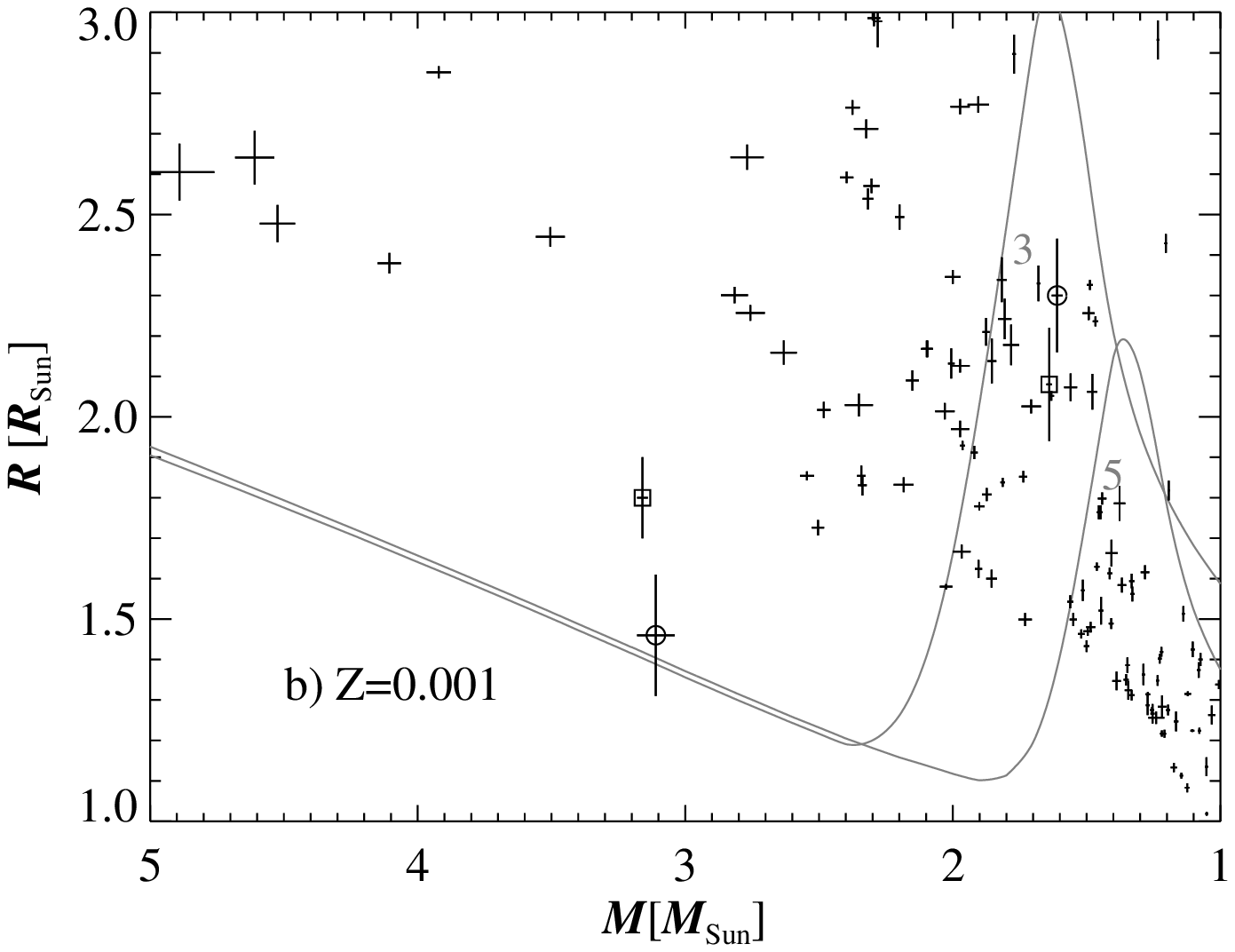}
 \includegraphics[width=8.2cm]{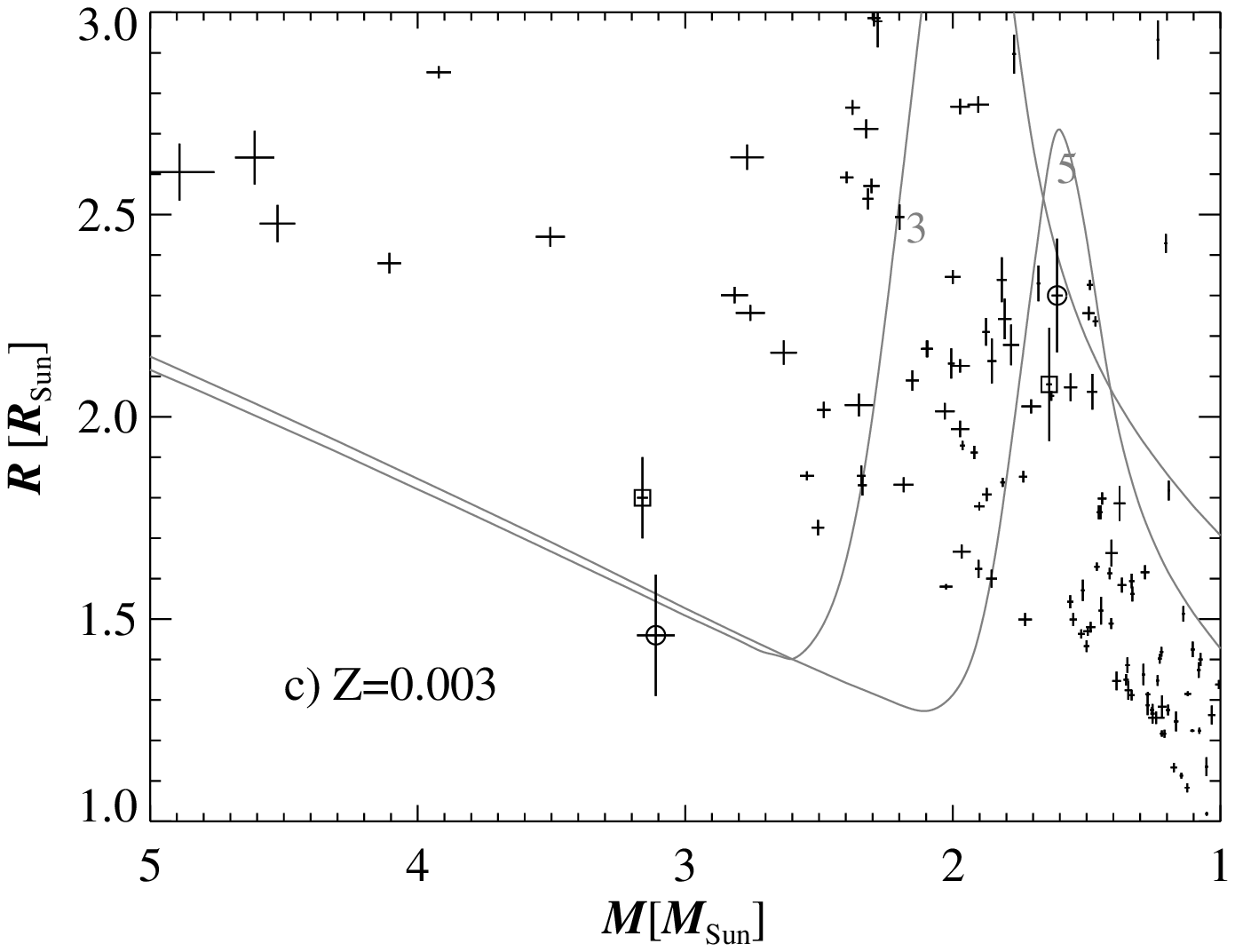}
% \vspace*{-1.0 cm}
 \caption{{\bf Mass-radius diagram}. The new measurements of TY\,CrA (circles) are compared to the previous 
 results obtained by \citet[][squares]{casey+98} and the data of other binaries compiled by
 \citet{torres+10}. The measurements are compared to theoretical isochrones for 3 and 5\,Myrs from
 \citet{tognelli+11} at different metallicity: {\bf a)} $Z$=0.0125, $Y$=0.255, $X_\mathrm{D}=2\cdot10^{-5}$, $\alpha=1.20$; {\bf b)} 
 $Z$=0.001, $Y$=0.232, $X_\mathrm{D}=4\cdot10^{-5}$, $\alpha=1.20$; {\bf c)}
 $Z$=0.003, $Y$=0.236, $X_\mathrm{D}=4\cdot10^{-5}$, $\alpha=1.20$.
}
   \label{fig:mr}
\end{center}
\end{figure}

Masses and radii are the most fundamental parameters derived from the observations. They are hardly affected by uncertainties 
in the adopted effective temperature of the primary component. We varied the primary temperature in the light curve analysis 
from 10,000\,K to 14,000\,K in order to see whether there is any effect on the resulting masses and radii. While luminosity 
varies drastically from 30 to more than $70\,\Lsun$, the changes in masses and radii remain well below the formal error bars. 
In other words, the derived values of mass and radius are very robust w.r.t. an uncertain primary $\Teff$.
 
The radii are indicative of the evolutionary state of both components. The mass-radius diagram in Fig.~\ref{fig:mr} compares 
the loci of both components to previous assessments by \citet{casey+98} and precise data of other binaries 
\citep{torres+10}. In addition, PISA isochrones of \citet{tognelli+11} of different age and metallicity are overplotted. 
\citet{casey+98} have already noticed that TY\,CrA has reached the main sequence while the secondary still resides in 
the pre-main sequence phase.

Not only the secondary but also the primary is an important object since it is located in an otherwise empty region of 
the mass-radius diagram at small radii. Our value of the primary radius is even lower than the value given by \citet{casey+98}. 
The comparison to the \citet{torres+10} data shows that the primary component seems considerably undersized. The primary 
only agrees with the models when adopting a metallicity of $Z$=0.003 or less (Figs.~\ref{fig:mr}b and \ref{fig:mr}c). For 
comparison, Fig.~\ref{fig:mr}a shows the $Z$=0.0125 isochrone according to the metallicity adopted by \citet{gennaro+12}.
\citet{casey+98} compared to models at even higher metallicity, Sun-like and Hyades-like.

The secondary agrees with a very young age of some 3-5\,Myrs for any set of isochrones, 
in accordance with previous estimates by \citet[][$\approx3\,$Myrs]{casey+98} and \citet[][$\approx4-5\,$Myrs]{gennaro+12}. 

\subsection{HR diagram}

\begin{figure}
% \vspace*{-2.0 cm}
\begin{center}
 \includegraphics[width=8.2cm]{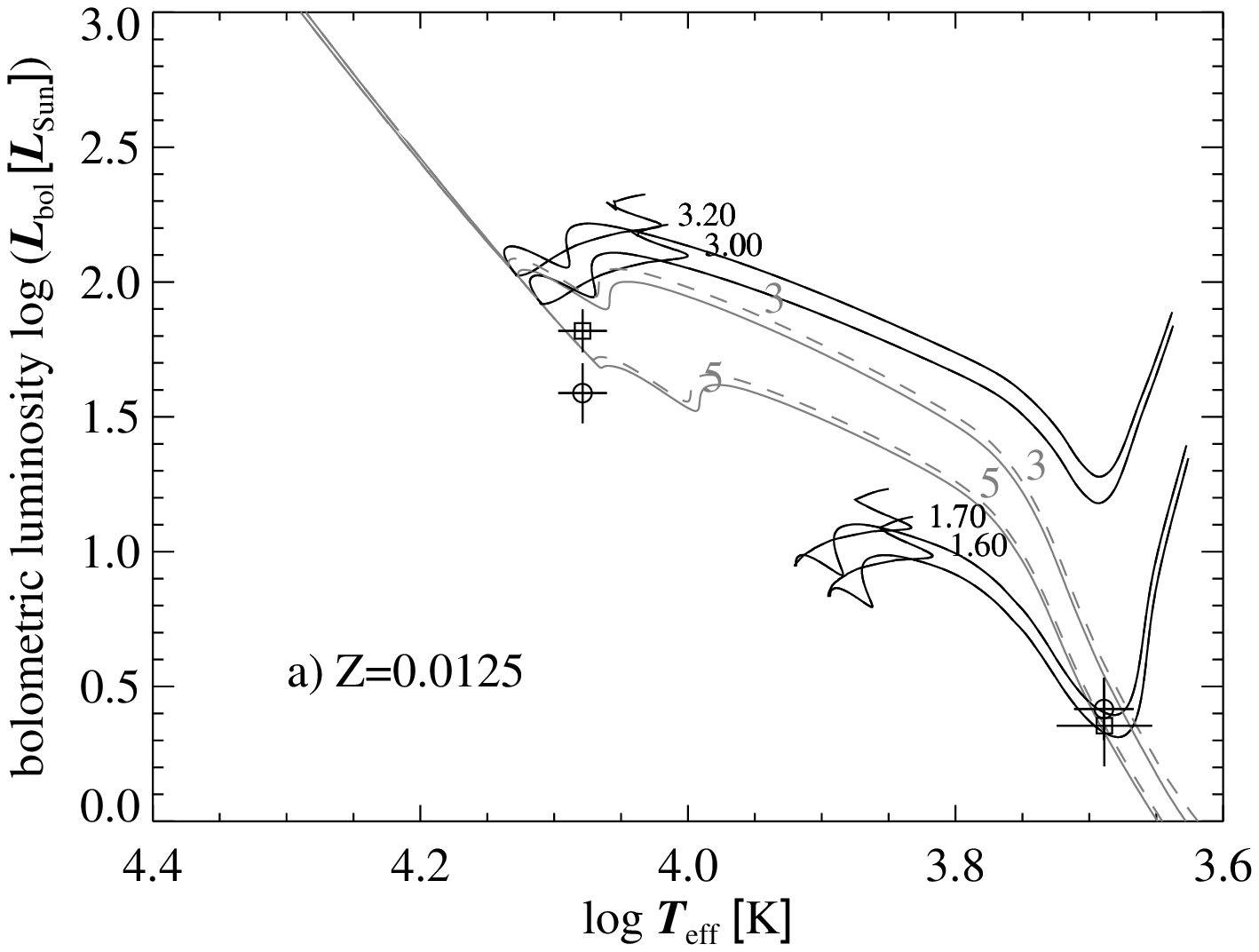}
 \includegraphics[width=8.2cm]{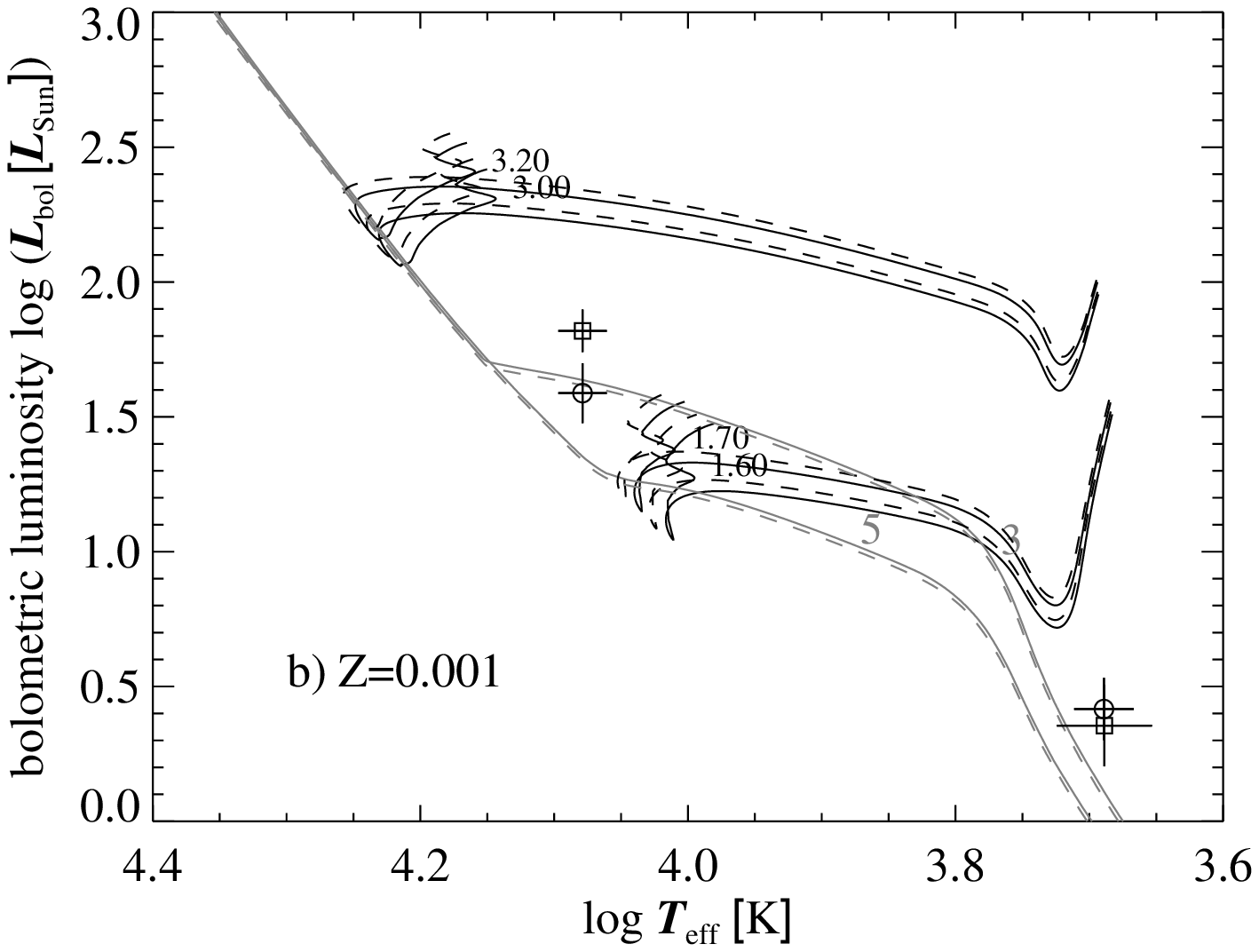}
 \includegraphics[width=8.2cm]{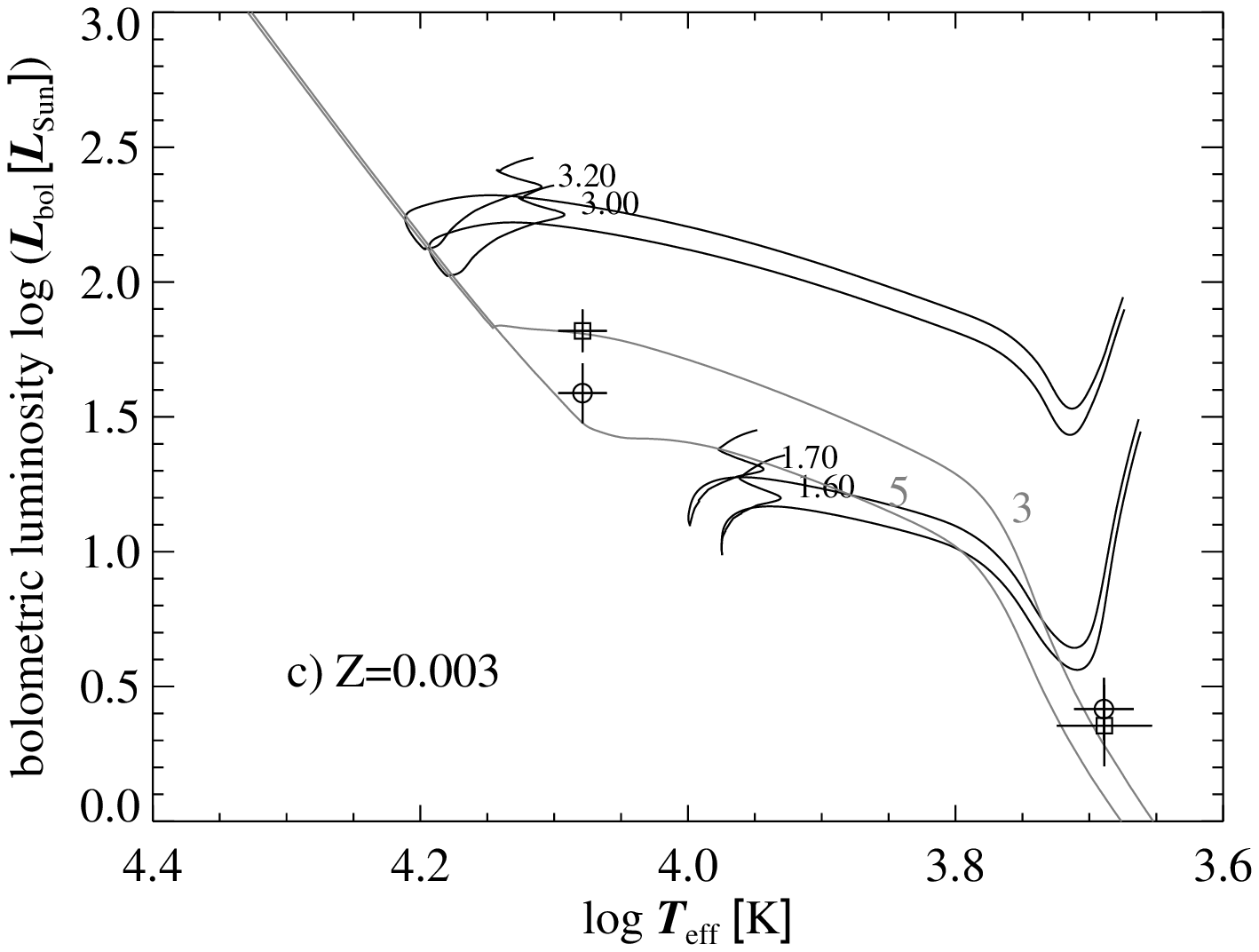}
% \vspace*{-1.0 cm}
 \caption{{\bf HR diagram.} The new measurements of TY\,CrA (circles) are compared to the previous results obtained by 
 \citet[][squares]{casey+98}. The measurements are compared to PISA models \citep{tognelli+11} at different metallicity. 
 Theoretical isochrones for 3\,Myrs and 5\,Myrs are displayed as well as mass tracks for 1.6, 1.7, 3.0, and 
 3.2\,$\Msun$: {\bf a)} $Z$=0.0125, $Y$=0.255, $X_\mathrm{D}=2\cdot10^{-5} \mathrm{(solid)} / 4\cdot10^{-5}$ (dashed), $\alpha=1.20$ (sets 
 with different $X_\mathrm{D}$ are used to enable comparison with \citet{gennaro+12}, and to show that different 
 $X_\mathrm{D}$ has relatively small effect); {\bf b)} $Z$=0.001, $Y$=0.232 (solid) / 0.251 (dashed), 
 $X_\mathrm{D}=4\cdot10^{-5}$, $\alpha=1.20$ (sets with different $Y$ are used to enable comparison with other 
 metallicities, and to show that different $Y$ has relatively small effect); {\bf c)}
 $Z$=0.003, $Y$=0.236, $X_\mathrm{D}=4\cdot10^{-5}$, $\alpha=1.20$.
}
   \label{fig:hrd}
\end{center}
\end{figure}

The evolution of the radiative quantities luminosity and effective temperature is studied in the HR diagram 
(Fig.~\ref{fig:hrd}). Again, the new measurements are compared to the previous location in the HR diagram according 
to \citet{casey+98} and to PISA models, now including stellar tracks. For clarity and since we do not gain additional 
insight, the \citet{torres+10} data is not shown in Fig.~\ref{fig:hrd}. 

Figure~\ref{fig:hrd}a aims at reproducing fig.~8 of \citet{gennaro+12} who fitted the location in the HR 
diagram based on the \citet{casey+98} data. However, Fig.~\ref{fig:hrd}a looks more complicated since the 
very same tracks and isochrones are not available to us. Therefore, the two mass tracks closest to the 
measured masses are shown for each component of TY\,CrA. In addition, models for different deuterium abundances are 
shown in order to be able to compare with Figs.~\ref{fig:hrd}b and ~\ref{fig:hrd}c. However, the effect of deuterium 
abundance is small compared to the variations due to age, mass, and metallicity. Obviously, the tracks and isochrones
are at variance with the new primary data. An isochrone of older age might still be consistent with the primary, but 
would disagree with the secondary. Furthermore, the discrepancy with the primary in the mass-radius diagram 
(Fig.~\ref{fig:mr}a) would not be improved.

In contrast to the mass-radius diagram, the overall situation in the HR diagram does not improve when choosing 
low-metallicity models. At least, choices of low metallicity allow for agreement of the primary with isochrones 
of 3\,Myrs (Fig.~\ref{fig:hrd}b)\footnote{In Fig.~\ref{fig:hrd}b, models for two different helium abundances 
are shown to allow for a better comparison with Figs.~\ref{fig:hrd}a and \ref{fig:hrd}c. Again, the effect is 
comparably small, in a way similar to the effect of varying deuterium abundance.} and 5\,Myrs (Fig.~\ref{fig:hrd}c), 
respectively. The disagreement with the mass tracks, however, is exacerbated. Moreover, the low-metallicity tracks 
disagree with the secondary in the HR diagram. For $Z$=0.003 (Fig.~\ref{fig:hrd}c), the evolutionary tracks 
agree at best marginally with the new secondary parameters. Moreover, no isochrone can be found which matches both 
components. In summary, low metallicity does not explain the small primary radius, even though this is suggested 
by the mass-radius diagram.

It is interesting to note, that the models of $Z$=0.0125 might agree with the observations in the HR diagram when 
adopting a different value of the primary effective temperature which is an input parameter in the LC 
analysis. While there is no significant effect on the radius measurements, the luminosities derived vary substantially. 
Nevertheless, the discrepancies in the mass-radius diagram will persist.

\section{Summary}

We present new optical and NIR photometry of TY\,CrA taken in 2006-2011. The
$BVRI$ photometry taken complements the Str\"omgren photometry studied by 
\citet{casey+98}. For the first time, we present NIR light curves of the secondary
minimum taken in the $J$ and $H$ bands. In contrast to the optical light curves the secondary minimum is much better 
revealed in the NIR and is more than $0.1\,$mag deep

We compiled all minima times available to us so that the O-C diagram of TY\,CrA now spans over two decades. 
The light-time effect of the spectroscopic tertiary is too small to explain the changes in the O-C diagram. 
There are probably other effects complicating the analysis like shifts of the observed minima due to photometric 
out-of-eclipse variability. We find, that RV measurements are more suitable to solve the tertiary orbit. 
One requires high-resolution high S/N spectroscopy with an echelle spectrograph at a 2-m class telescope 
covering at least one orbital period of the unresolved triple system (270 days most probably). 
The Fourier-domain technique \citep[see][]{hadrava+95} then can be used to disentangle the 
component's spectra for further analysis ($\Teff$, $\log g$, $\log$ [m/X]) and to improve 
the spectroscopic  orbits. The orientation of the components' rotational axes could be found analyzing 
the McLaughlin rotational effect during eclipses of the inner pair. 
Long-baseline interferometry would help to determine orbital parameters of the outer orbit of the 
triple and to determine the distance to the system (comparing interferometric and spectroscopic 
semi-major axis of the inner triple system) and the total mass of the triple.

For the first time, we present a multi-band solution of the light curve of the inner eclipsing 
binary in both optical and near-$IR$ bands. The near-$IR$ data requires to account for a third light 
contribution and thus also allows us to constrain the colours of the third light. 
The third light is composed of at least three sources which are the reflection nebula, 
the spectroscopic tertiary, and the visual fourth component.

Only the radius and the luminosity of the primary do not agree with the values published 
by \citet{casey+98}. Agreement is not improved by restricting the analysis to the optical data. 
While the primary luminosities could be reconciled by adjusting the effective temperature of the primary, 
the derived primary radius is very robust and would not change.

Indeed, the absolute parameters of TY\,CrA do not agree with expectations. The primary component seems undersized when 
compared to other stars of same mass and to recent PISA evolutionary tracks and isochrones. The HR diagram 
shows that low metallicity cannot explain the small radius of the primary. As speculated by \citet{casey+98}, 
the subsynchronous rotation of the primary and effects on its internal structure might play a role.

A very interesting question is whether the fourth component is actually bound 
to the TY\,CrA system. In order to decide whether the fourth component is bound, 
there is need for single-epoch AO observation of the system to establish the physical 
bond of the fourth visual component 0\farcs3 away in 2002. If it
was a background star, the proper motion of TY~CrA (see Table~\ref{comparison_tab}) 
would make the visual companion unresolved around 2012. Because of the high total 
mass of the whole system the orbital period of the outermost visual orbit is probably rather short: 
assuming a distance of $d$ = 130 pc, the 0\farcs294 separation observed in 2002 as the semi-major 
axis and the total mass of the system as 7 $M_\odot$, we get $P_{1234}$ = 89.3 years.

%Improving the system's parameters would require:

% \item high-precision satellite photometry to improve photometric elements (component radii and inclination angle).

% Note maybe it should be quantified

% In spite of poor weather conditions during the NIR observing runs the ...
 
% Q: what is the nature of the reflection nebula: is it just the scatter on free electrons, do we see some emission
% lines ?

\section*{Acknowledgments}

We wish to thank Vera Hoffmeister for taking observations with VYSOS6
and Fred Walter for crucial ideas on the observing strategy and for taking
observations with ANDICAM. We are indebted to Torsten Sch\"oning who strongly 
contributed in the early beginning of the project. We wish to thank Eugenio Schisano
and Viki Joergens for their support in the preparation of observations and
helpful suggestions. We acknowledge help by Markus Mugrauer and Ronny Errmann in the 
reduction of SOFI and REMIR data. 

This work has been funded by the project DFG AM 158/3-1. M.V. and T.P. would like to thank the project VEGA 2/0094/11 
and the project APVV-0158-11. M.A. was supported by a graduate scholarship of the Cusanuswerk, one of
the national student elite programs of Germany and by a scholarship (reference SFRH/BPD/26817/2006) granted by 
the \textit{Funda\c{c}\~ao  para a Ci\^encia e a Tecnologia} (FCT), Portugal. M.A. acknowledges    
research funding granted by the \textit{Deutsche Forschungsgemeinschaft} (DFG) under the project RE 1664/4-1.
M.A. further acknowledges support by \textit{Deutsches Zentrum f\"ur Luft- und Raumfahrt} DLR under the project 
50OW0204. This publication is supported as a project of the Nordrhein-Westf\"alische Akademie 
der Wissenschaften und der K\"unste in the framework of the academy program by the Federal Republic of Germany 
and the state Nordrhein-Westfalen. M.V., T.P. and R.N. acknowledge support from the EU in the FP6 MC ToK project
MTKD-CT-2006-042514. This research has made use of the SIMBAD database, operated at CDS, Strasbourg, France, and NASA’s    
Astrophysics Data System Bibliographic Services. 

{}

\label{lastpage}
\end{document}